# Model-Free Adaptive Control Compensated with Disturbance

Feilong Zhang

***Abstract*—In this paper, we restudy how to modify the model-free adaptive control (MFAC) to reject the disturbances in both single-input single-output (SISO) systems and multiple-input multiple-output (MIMO) systems. This research endeavor is intended to pave the way for future developments in the nonlinear theory related to this practical controller. To accurately describe the nonlinear system model at each time, we first compensate for the equivalent dynamic linearization model (EDLM) with disturbances and prove it according to the definition of differentiability and the Taylor series. Based on the modified EDLM, we then redesign MFAC compensated with disturbances and analyze the performance of the nonlinear system using the closed-loop system equation at each time. This is all possible because some nonlinear system functions can be accurately described by the EDLM compensated with disturbances, according to the Taylor series or definition of differentiability. Finally, several examples are provided to validate the theorem.**

## I. Introduction

Considerable work about MFAC has been published during the past decade. The controller design relies on a process model called the equivalent dynamic linearization model, whose coefficients constitute the pseudo-gradient (PG) vector for the SISO system or the pseudo-Jacobian matrix (PJM) for the MIMO system. The time-varying PG vector or PJM was typically estimated online using the projection estimation/least square method. This kind of process model has been classified into three forms: compact form (CF) EDLM ($L_y$=0, $L_u$=1), partial form (PF) EDLM ($L_y$=0, $L_u\geq$1), and full form (FF) EDLM ($L_y\geq$0, $L_u\geq$1). The MFAC controller is obtained by optimizing the quadratic index function combined with the EDLM. Since MFAC based on FF-EDLM incorporates those designed based on two other forms [1]-[3], this paper focuses solely on designing and analyzing the MFAC based on FF-EDLM in detail.

If we aim to study the nature of this kind of adaptive control or judge the correctness of some related theorems, we should start by analyzing this control method in the same way as [4]-[6] begin with deterministic and linear systems. One undeniable reason is that the nonlinear systems incorporate the linear systems [7]. The theoretical basis of the current MFAC design is to describe the nonlinear system model by using EDLM at any point, according to the principle of Cauchy mean value theorem [1]-[3]. More precisely, EDLM refers to the process of locally linearizing the nonlinear system function, which naturally characterizes the designed controller with a linear incremental form. Consequently, the design process of the MFAC controller is fundamentally similar to that of linear controllers based on linear systems in [4]-[6]. In addition, the adaptability of MFAC to the nonlinearity and uncertainty is achieved by combining the designed controller with the online parameter estimator, according to the certainty-equivalent principle [4]. Furthermore, [8] points out that the essence of adaptability introduced by online estimated algorithms is to provide a more accurate reflection of the system model rather than being model-free. Therefore, our studies [8]-[11] on MFAC primarily focus on both deterministic linear systems and deterministic nonlinear systems. Additionally, the controller coefficients are designed according to the actual system model for accurately understanding the nature of this kind of controller. These may help to develop this topic.

In practical situations, the input and output signals may be affected by external disturbances. To design a disturbance rejection controller, [12] multiplies the compact form MFAC by an attenuation coefficient and proves a noticeable conclusion that the tracking error of the system finally converges to zero when $\lambda$ is large enough. Similarly, [13] compensates the MFAC with the disturbance item and introduces an observer to estimate the disturbance for SISO systems. However, if we want to clearly study the basic principle of this kind of MFAC compensated with the estimated disturbance, we should not add the observer at the beginning and instead use the actual disturbance, according to the single variable principle (single variable method). Along with this, some deficiencies in [13] naturally appear in our linear system example. In addition, [14] applies the MFAC compensated with the disturbance to the MIMO nonlinear heterogeneous multiagent systems. To expose its fundamental errors, we should initially focus on a single separate agent and subsequently find that the proposed controller in [14] can be regarded as the multivariable MFAC compensated with the disturbance proposed in this paper. [14] proves a theorem that the tracking errors of the system are bounded when $\lambda$ is large enough. However, its theorem is entirely inconsistent with the conclusion in this paper. Truth is the unity of universality and particularity. Many claimed conclusions regarding MFAC compensated with disturbance in current works are not valid for simple and easy linear systems, let alone for nonlinear systems.

Manuscript received May 25, 2021. This work was supported in part by the xxxxx.

Feilong Zhang is with the State Key Laboratory of Robotics, Shenyang Institute of Automation, Chinese Academy of Sciences, Shenyang 110016, China (e-mail: zhangfeiong@sia.cn).

This paper introduces the disturbance item into FF-EDLM and proves this modified EDLM according to the definition of differentiability and Taylor series, rather than relying on the Cauchy mean value theorem. Simultaneously, the Taylor series also provides an essential method for calculating the coefficients of the EDLM, which constitute the PG vector or the PJM. Afterward, we combine the quadratic index function and the modified EDLM and solve for its optimal solution to obtain the MFAC compensated with disturbance. In addition, it might be the first time we analyze the discrete-time nonlinear system performance by an easy yet extraordinary method, i.e., closed-loop system equation at each time. This is all possible because some nonlinear system functions can be accurately described by the EDLM compensated with disturbance. Moreover, this method also provides the disturbance-to-output transfer function, which fundamentally clarifies some implausible relationships between the disturbance and the key parameter $\lambda$ in some current works. To validate our findings, we conducted several illustrative examples.

This paper is organized as follows. Section II presents the modified EDLM, and the MFAC compensated with disturbance for SISO nonlinear systems. Then the system performance is analyzed through the closed-loop system function, and the simulations are given to validate our viewpoints in both the nonlinear and linear systems. Similar to Section II, Section III introduces the modified EDLM and the MFAC, which are compensated with disturbance for MIMO nonlinear systems. Subsequently, the system performance is analyzed through the closed-loop system function, and two simulations verify our viewpoints. Section IV presents the conclusions. Finally, the proof of the modified EDLM for SISO systems is provided in the Appendix.

## II. Equivalent Dynamic Linearization Model and Design of Model-Free Adaptive Control for SISO Systems

In part A of this section, we modify and prove the EDLM with disturbance for SISO systems. Part B presents the design of the MFAC compensated with disturbance and the system performance analysis. Part C gives simulations for the verification of the theory.

### A. *Equivalent Dynamic Linearization Model with Disturbance for SISO Systems*

The discrete-time SISO system is considered as

$$y(k+1)=f(y(k),\cdots,y(k-n_y),u(k),\cdots,u(k-n_u))+w(k+1) \tag{1}$$

where $f(\cdots)\in R$ represents a linear or a nonlinear differentiable function. $w(k)$, $u(k)$ and $y(k)$ represent the disturbance, control input and output of the system at time $k$, respectively. And $n_y+1$ ($n_u+1$)$\in$Z is the order of the system output (input) [4]-[6], [15]. Let

$$\boldsymbol{\varphi}(k)=[y(k),\cdots,y(k-n_y),u(k),\cdots,u(k-n_u)]^T \tag{2}$$

Then (1) can be rewritten as

$$y(k+1)=f(\boldsymbol{\varphi}(k))+w(k+1) \tag{3}$$

*Theorem 1*: Given that $\Delta\boldsymbol{H}(k)\neq 0$, a time-varying vector $\boldsymbol{\phi}_L(k)$ named PG vector must exist and the system (1) can be rewritten into the following EDLM compensated with disturbance:

$$\Delta y(k+1)=\boldsymbol{\phi}_L^T(k)\Delta\boldsymbol{H}(k)+\Delta w(k+1) \tag{4}$$

where

$$\boldsymbol{\phi}_L(k)=\begin{bmatrix}\boldsymbol{\phi}_{Ly}(k)\\ \boldsymbol{\phi}_{Lu}(k)\end{bmatrix}=[\phi_1(k),\cdots,\phi_{Ly}(k),\phi_{Ly+1}(k),\cdots,\phi_{Ly+Lu}(k)]^T,$$

and

$$\Delta\boldsymbol{H}(k)=\begin{bmatrix}\Delta\boldsymbol{Y}_{Ly}(k)\\ \Delta\boldsymbol{U}_{Lu}(k)\end{bmatrix}=[\Delta y(k),\cdots,\Delta y(k-L_y+1),\ \Delta u(k),\cdots,\Delta u(k-L_u+1)]^T.$$

Two integers $0\leq L_y$, $1\leq L_u$ are called pseudo orders of the system.

*Proof*: please refer to Appendix.

We define $\boldsymbol{\phi}_{Ly}(z^{-1})=\phi_1(k)+\cdots+\phi_{Ly}(k)z^{-Ly+1}$, $\boldsymbol{\phi}_{Lu}(z^{-1})=\phi_{Ly+1}(k)+\cdots+\phi_{Ly+Lu}(k)z^{-Lu+1}$, and $z^{-1}$ is the backward-shift operator.

### B. *Design of Model-Free Adaptive Control for SISO Systems*

(4) is easily rewritten into (5).

$$y(k+1)=y(k)+\boldsymbol{\phi}_L^T(k)\Delta\boldsymbol{H}(k)+\Delta w(k+1) \tag{5}$$

The object is to design a controller that optimizes the cost function:

$$J=\left|y^*(k+1)-y(k+1)\right|^2+\lambda\left|\Delta u(k)\right|^2=\min imum \tag{6}$$

where $y^*(k+1)$ is the desired system output and $\lambda$ is the weighted constant. Substituting (5) into (6) and solving $\partial J/\partial\Delta u(k)=0$ yield

$$\begin{aligned}\Delta u(k)=\frac{\phi_{Ly+1}(k)}{\lambda+\phi_{Ly+1}^2(k)}[&y^*(k+1)-y(k)-\sum_{i=1}^{Ly}\phi_i(k)\Delta y(k-i+1)\\&-\sum_{i=Ly+2}^{Ly+Lu}\phi_i(k)\Delta u(k+L_y-i+1)-\Delta w(k+1)]\end{aligned} \tag{7}$$

Considering that the disturbance $w(k)$ may not be acquired directly, we replace it by $\hat{w}(k)$ which represents the measurement or the estimation of the disturbance. Consequently, (7) is rewritten as

$$\begin{aligned}\Delta u(k)=\frac{\phi_{Ly+1}(k)}{\lambda+\phi_{Ly+1}^2(k)}[&y^*(k+1)-y(k)-\sum_{i=1}^{Ly}\phi_i(k)\Delta y(k-i+1)\\&-\sum_{i=Ly+2}^{Ly+Lu}\phi_i(k)\Delta u(k+L_y-i+1)-\Delta\hat{w}(k+1)]\end{aligned} \tag{8}$$

Form (4) and (8), we have the closed-loop system equation (9) at the time $k$:

$$\begin{aligned}&\left[\lambda(1-z^{-1})\left[1-z^{-1}\boldsymbol{\phi}_{Ly}(z^{-1})\right]+\phi_{Ly+1}(k)\boldsymbol{\phi}_{Lu}(z^{-1})\right]y(k)\\&=\phi_{Ly+1}(k)\boldsymbol{\phi}_{Lu}(z^{-1})y^*(k)+[\lambda+\phi_{Ly+1}(k)\boldsymbol{\phi}_{Lu}(z^{-1})]\Delta w(k)\\&\quad-\phi_{Ly+1}(k)\boldsymbol{\phi}_{Lu}(z^{-1})\Delta\hat{w}(k)\end{aligned} \tag{9}$$

We may place the closed-loop poles in the unit circle by tuning $\lambda$ to obtain the inequality (10).

$$T(z^{-1}) = \lambda(1-z^{-1})\left[1-z^{-1}\boldsymbol{\phi}_{Ly}(z^{-1})\right]+\phi_{Ly+1}(k)\boldsymbol{\phi}_{Lu}(z^{-1}), \quad |z|>1 \tag{10}$$

If $\Delta\hat{w}(k) = \Delta w(k)$, the disturbance-to-output transfer function is written as

$$G_w(z^{-1}) = \frac{\lambda(1-z^{-1})}{\lambda(1-z^{-1})\left[1-z^{-1}\boldsymbol{\phi}_{Ly}(z^{-1})\right]+\phi_{Ly+1}(k)\boldsymbol{\phi}_{Lu}(z^{-1})} \tag{11}$$

Additionally, when $\lambda$=0, we can rewrite the closed-loop system equation (9) into

$$\begin{aligned}&\left[\lambda(1-z^{-1})\left[1-z^{-1}\boldsymbol{\phi}_{Ly}(z^{-1})\right]+\phi_{Ly+1}(k)\boldsymbol{\phi}_{Lu}(z^{-1})\right]y(k)\\ &=\phi_{Ly+1}(k)\boldsymbol{\phi}_{Lu}(z^{-1})y^{*}(k)\end{aligned} \tag{12}$$

which indicates that the influence of disturbance $w(k)$ is theoretically removed when (10) is satisfied.

On the other hand, many estimators are designed to estimate the unknown disturbance, and we have developed a simple one as follows:

$$\hat{w}(k+1) = \hat{w}(k) - L\left[\hat{w}(k) - w(k)\right] \tag{13}$$

$$w(k) = f(\boldsymbol{\varphi}(k-1)) - y(k) \tag{14}$$

or

$$\Delta\hat{w}(k+1) = \Delta\hat{w}(k) - L\left[\Delta\hat{w}(k) - \Delta w(k)\right] \tag{15}$$

$$\Delta w(k) = \Delta y(k) - \boldsymbol{\phi}_L^T(k-1)\Delta\boldsymbol{H}(k-1) \tag{16}$$

Then we have the relationship between the disturbance and its estimation at the time $k$:

$$\hat{w}(k+1) = \frac{zL}{z+(L-1)}w(k) \tag{17}$$

From (17), we know that the estimator is stable when $L \in [0, 2]$. When $L$=1 or $z\rightarrow 1$, (17) becomes to $\hat{w}(k+1) = z^{-1}w(k+1)$.

*C. Simulations:*

*Example 1*: In this example, the following discrete-time SISO nonlinear system is considered.

$$y(k+1) = -y^2(k) + u(k) + 0.2u^2(k-1) + w(k+1) \tag{18}$$

where the disturbance is

$$w(k+1) = 0.5\sin(k/40) + 0.5\cos(k/30) \tag{19}$$

The desired trajectory is

$$y^{*}(k) = 0.3\times(-1)^{round(k/50)}, \quad 1 \le k \le 700 \tag{20}$$

According to [4]-[6], the controller structure should be applied with $L_y$=$n_y$+1=1 and $L_u$=$n_u$+1=2. The elements in PG are calculated through $\hat{\phi}_2(k)=1$,

$$\hat{\phi}_1(k) = \sum_{i=1}^{2}\frac{1}{i!}\frac{\partial^i f(\boldsymbol{\varphi}(k-1))}{\partial y^i(k-1)}\Delta y^{i-1}(k) = -2y(k-1) - \Delta y(k)$$ and

$$\hat{\phi}_3(k) = \sum_{i=1}^{2}\frac{1}{i!}\frac{\partial^i f(\boldsymbol{\varphi}(k-1))}{\partial u^i(k-2)}\Delta u^{i-1}(k) = 0.2\times(2u(k-2)+\Delta u(k-1))$$

We estimate the disturbance using (13), (14) and choose $L$=1 for the optimal estimate performance. The outputs of system controlled by (8) with $\lambda$=0, $\lambda$=1.5 and $\lambda$=3 are shown in Fig. 1. Fig. 2 shows the corresponding control inputs. Fig. 3 illustrates the elements in the calculated PG vector for a specific value of $\lambda$=0. Fig. 4 presents the disturbance and its corresponding estimation.

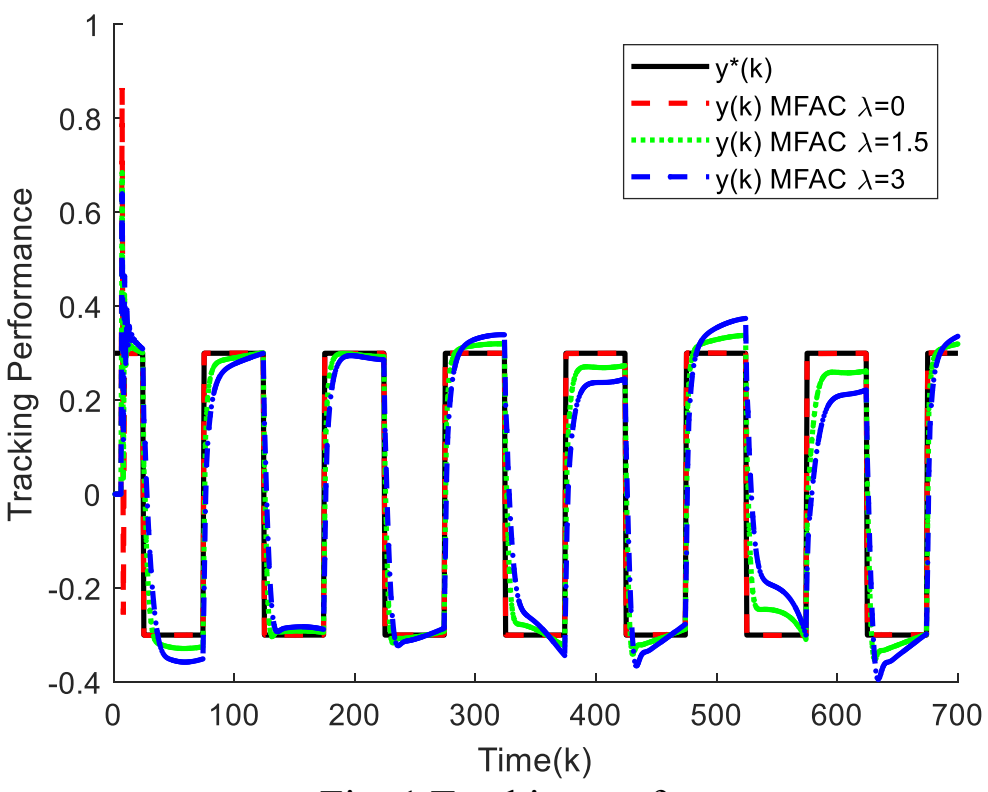

Fig. 1 Tracking performance

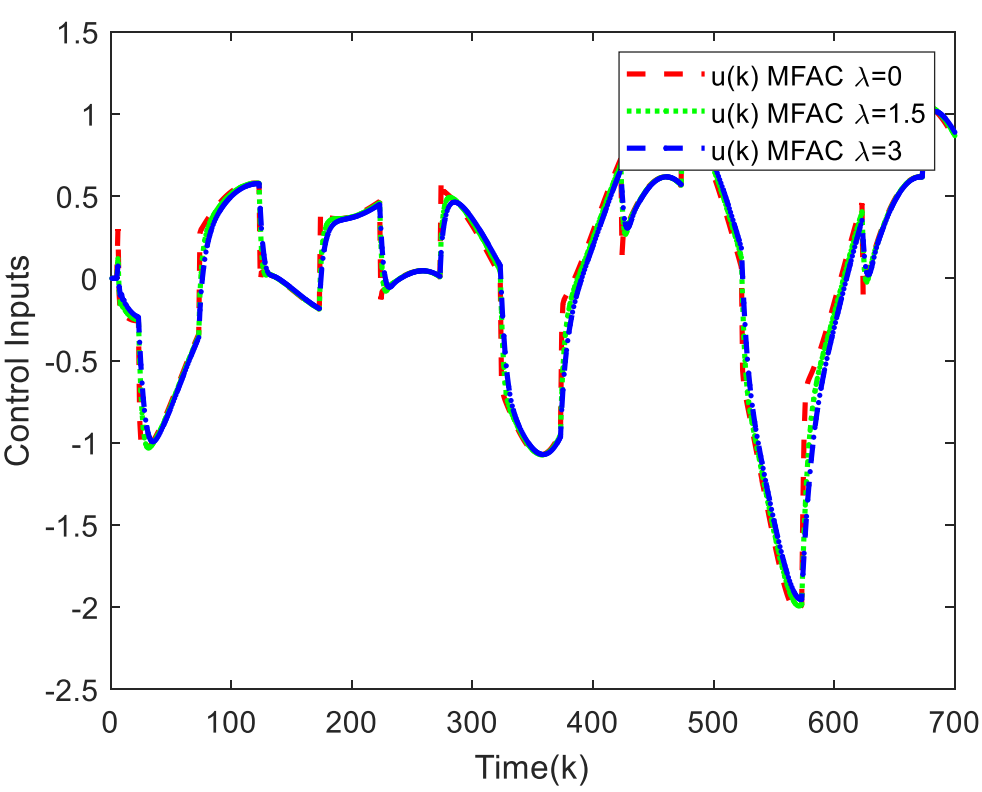

Fig. 2 Control input

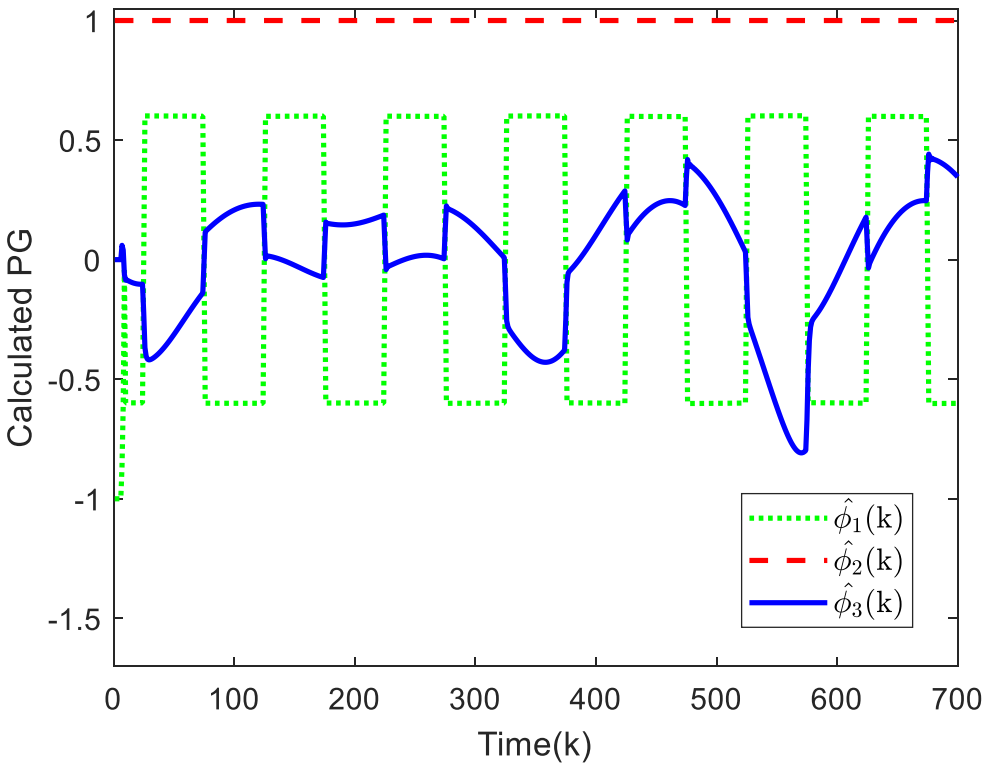

Fig. 3 Elements in calculated PG vector

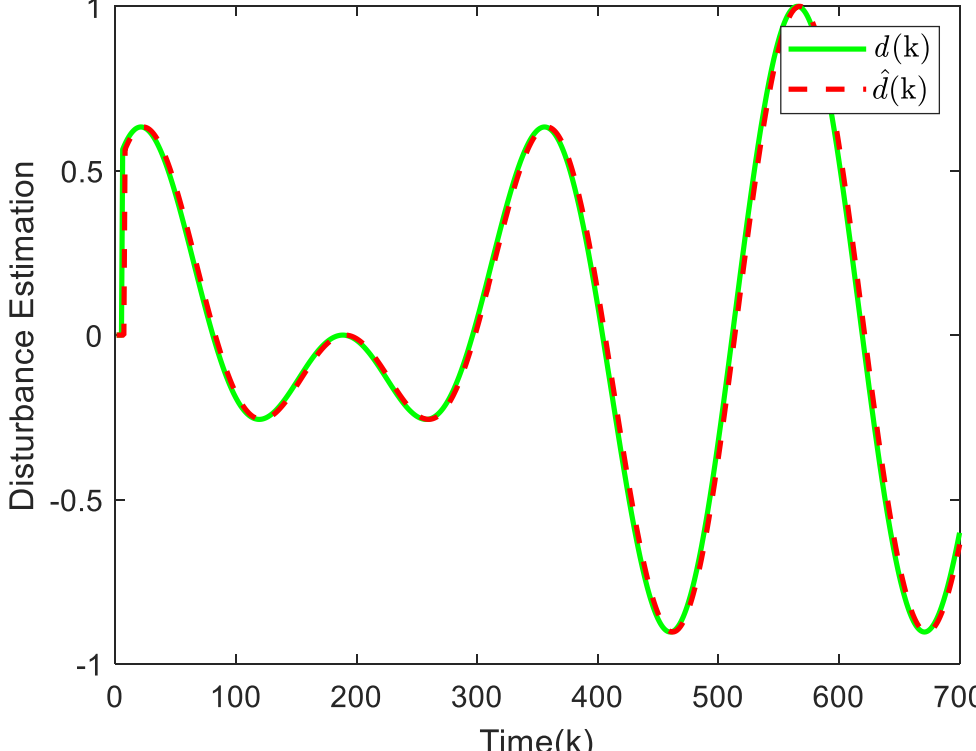

Fig. 4 Disturbance and its estimation

In Fig. 1, it is evident that the influence of the disturbance is almost removed when $\lambda$=0. Moreover, the influence of disturbance increases as $\lambda$ is raised. This observation contradicts certain theorems in current literature, such as [12]-[14]. Fig. 4 demonstrates that the estimation of the disturbance $\hat{d}(k)$ lags behind the actual disturbance $d(k)$ by one control period.

*Example 1.1*: If we change the model (18) into

$$y(k+1) = -y^2(k) + u(k) + w(k+1) \tag{21}$$

We assume that the disturbance is known for the studies on the nature of MFAC compensated with disturbance. The controller is designed in accordance with (7). When the disturbance is the unit speed signal $w(k+1)=k$, the application of the final value theorem yields the system output caused by disturbance as

$$e = \lim_{z\to 1}(1-z^{-1})\frac{\lambda(1-z^{-1})}{\lambda(1-z^{-1})\left[1-z^{-1}\boldsymbol{\phi}_{Ly}(z^{-1})\right]+\phi_{Ly+1}(k)\boldsymbol{\phi}_{Lu}(z^{-1})}Z(w(k)) \\ = \lambda \tag{22}$$

where $Z(*)$ represents $z$-transformation. The outputs of the system controlled by MFAC compensated with disturbance (7) with $\lambda$=0, $\lambda$=±0.1 and $\lambda$=0.2 are shown in Fig. 5.

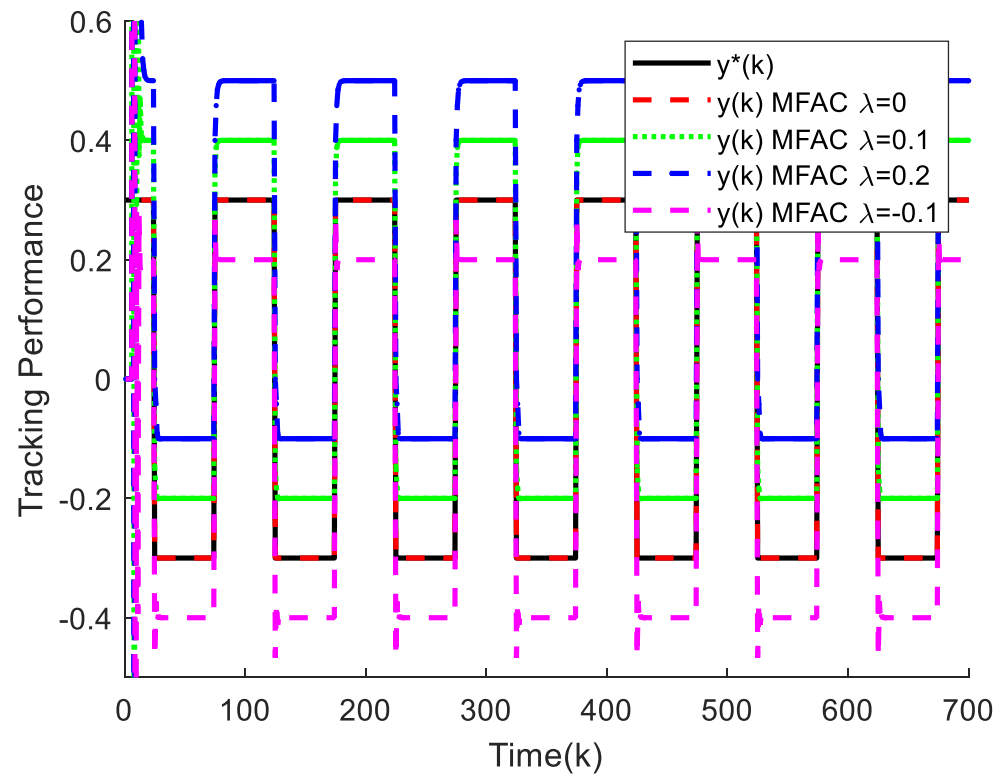

Fig. 5 Tracking performance

The steady-state error values are listed in Table I, based on the simulation results shown in Fig. 5. We can naturally conclude that the simulation results are consistent with our theorem.

TABLE I Measured tracking error $e(k)$ of the system

| $\lambda$ | 0 | ±0.1 | 0.2 |
|---|---|---|---|
| $e(40)=\cdots=e(700)$ | 0.000000 | ±0.100000 | 0.200000 |

*Example 2*: In this example, some problems in the Theorem 1 in [13] are shown. Herein, we will discuss the compact-form MFAC ($L_y=n_y+1=0$, $L_u=n_u+1=1$) compensated with disturbance for the following SISO stable linear system (23):

$$y(k+1) = f(u(k), w(k+1)) \\ = \phi(k)u(k) + w(k+1) = u(k) + 10\sin(k/10) \tag{23}$$

The desired trajectory is

$$y^*(k+1) = 5\times(-1)^{round(k/80)}, \quad 1\le k\le 400 \tag{24}$$

The initial values are $y(1)=\cdots=y(5)=u(1)=\cdots=u(5)=0$. The controller coefficient $\hat{\phi}(k)=1$ is set in accordance with the actual system model (23). If we want to study whether the controller plays a role in disturbance rejection and understand the underlying principle, we should not initially utilize the estimation of state variable (i.e., estimated disturbance) but the true one according to the single variable principle. We let $x_2(k)=\xi(k)=\Delta w(k+1)$ in [13], and then the controller (5) proposed in [13] is written into (25).

Fig. 6 shows the outputs of the system controlled by the MFAC compensated with disturbance (25) with $\lambda$=0, $\lambda$=1 and $\lambda$=10000.

$$\Delta u(k) = \frac{\phi(k)}{\lambda+\phi^2(k)}(y^*(k+1)-y(k)-\Delta w(k+1)) \\ = \frac{1}{\lambda+1}(y^*(k+1)-y(k)-10\sin(k/10)+10\sin((k-1)/10)) \tag{25}$$

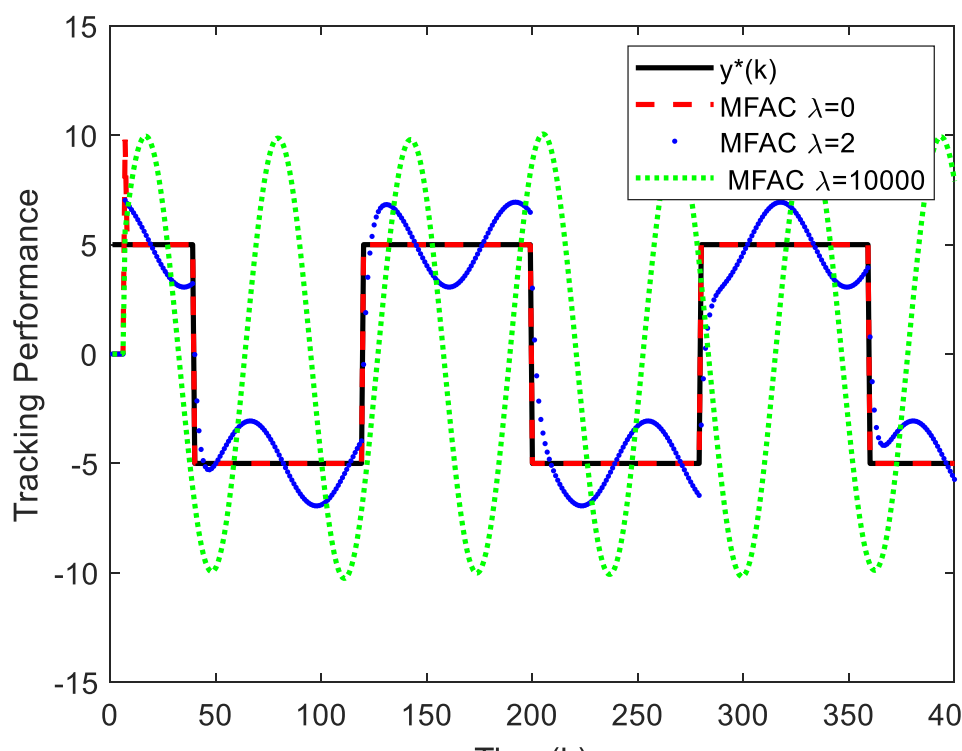

Fig. 6 Tracking performance

According to [13], we have $\left|\partial f/\partial u(k)\right| \le L_{n_y+2}=1$; $0<\omega\varepsilon\le\phi(k)\hat{\phi}(k)=1$; $b_{\hat{\phi}} > \max_{k\in[0,T]}\left|\hat{\phi}(k)\right|=1$ (we let $\vec{b}_{\hat{\phi}}=1.1$); $\xi(k)=\Delta w(k+1)=10\sin(k/10)-10\sin((k-1)/10)\le 1$, $\sup_k\left|\xi(k)\right|\le b_\xi$ (we let $b_\xi=1$); $b_{\hat{x}_2}=\sup_k\left|\hat{x}_2(k)\right|=\max_k\left|\Delta w(k+1)\right|=1$; $\rho=1$.

And [13] directly gives $0<c_3=\dfrac{\rho L_{ny+2}}{2\sqrt{\lambda}}<1$, $0<c_2=1-\dfrac{\rho\omega\varepsilon}{\lambda+b_{\hat{\phi}}}<1$.

According to Theorem 1 in [13], we have

$$\left|e(k)\right| < \frac{\delta_2}{1-c_2} = \frac{c_3 b_{\hat{x}_2}+b_\xi}{1-c_2} \le \left(\frac{\rho L_{ny+2}}{2\sqrt{\lambda}}+1\right)\Big/\frac{\rho\omega\varepsilon}{\lambda+b_{\hat{\phi}}} \\ = \frac{1+2\sqrt{\lambda}}{2\omega\varepsilon\sqrt{\lambda}}(\lambda+1.1) \tag{26}$$

From (26), we naturally deduce $\left|e(k)\right|<\infty$ as $\lambda$=0. On the contrary, the simulation shows $\left|e(k)\right|\to 0$ as $\lambda$=0.

Furthermore, the proof of Theorem 1 in [13] requires a sufficiently large $\lambda$ such that $c_2$ it is less than 1 on page 5. According to (26), it will deduce $\left|e(k)\right|<\infty$ as $\lambda=\infty$. However, this simulation shows that $e(k)=y^*(k)-w(k)$ as $\lambda=\infty$.

## III. Equivalent Dynamic Linearization Model and Design of Model-Free Adaptive Control for MIMO Systems

In part A of this section, we modify the EDLM with disturbance for MIMO systems and present its fundamental assumptions and theorem. Part B presents the design of MFAC compensated with disturbance and the system performance analysis. Part C gives the simulations to verify our viewpoints.

### *A. Equivalent Dynamic Linearization Model with Disturbance for MIMO systems*

The discrete-time MIMO nonlinear system is considered as

$$\boldsymbol{y}(k+1)=\boldsymbol{f}(\boldsymbol{y}(k),\cdots,\boldsymbol{y}(k-n_y),\boldsymbol{u}(k),\cdots,\boldsymbol{u}(k-n_u))+\boldsymbol{w}(k+1) \tag{27}$$

Define

$$\boldsymbol{\varphi}(k)=[\boldsymbol{y}(k),\cdots,\boldsymbol{y}(k-n_y),\boldsymbol{u}(k),\cdots,\boldsymbol{u}(k-n_u)] \tag{28}$$

and then (27) is rewritten as

$$\boldsymbol{y}(k+1)=\boldsymbol{f}(\boldsymbol{\varphi}(k))+\boldsymbol{w}(k+1) \tag{29}$$

where $\boldsymbol{f}(\cdots)=[f_1(\cdots),\cdots, f_{My}(\cdots)]^{\mathrm{T}}$ is the nonlinear vector-valued differentiable function. According to [4]-[6], $n_y+1$, $n_u+1\in \mathrm{Z}$ are the orders of output vector $\boldsymbol{y}(k)$, and input vector $\boldsymbol{u}(k)$ of the system at time $k$, respectively. $\boldsymbol{w}(k)$ represents the disturbance vector. The dimensions of $\boldsymbol{y}(k)$ and $\boldsymbol{w}(k+1)$ are both $M_y$ and the dimension of $\boldsymbol{u}(k)$ is $M_u$ ($M_u\geq M_y$).

*Theorem 2*: If $\Delta\boldsymbol{H}(k)\neq 0$, $0\leq L_y$, $1\leq L_u$, there exists a pseudo-Jacobian matrix $\boldsymbol{\phi}_L^T(k)$ and (27) can be transformed into

$$\Delta\boldsymbol{y}(k+1)=\boldsymbol{\phi}_L^T(k)\Delta\boldsymbol{H}(k)+\Delta\boldsymbol{w}(k+1) \tag{30}$$

where

$\boldsymbol{\phi}_L^T(k)=[\boldsymbol{\phi}_{Ly}^T(k),\boldsymbol{\phi}_{Lu}^T]$, $\boldsymbol{\phi}_{Ly}^T(k)=[\boldsymbol{\Phi}_1(k),\cdots,\boldsymbol{\Phi}_{Ly}(k)]_{My\times(Ly\bullet My)}$;

$\boldsymbol{\phi}_{Lu}^T(k)=[\boldsymbol{\Phi}_{Ly+1}(k),\cdots,\boldsymbol{\Phi}_{Ly+Lu}(k)]_{My\times(Lu\bullet Mu)}$;

$\boldsymbol{\Phi}_i(k)\in\boldsymbol{R}^{My\times My}$ ($i=1,\cdots,L_y$); $\boldsymbol{\Phi}_i(k)\in\boldsymbol{R}^{My\times Mu}$ ($i=L_y+1,\cdots,L_y+L_u$);

$\Delta\boldsymbol{H}(k)=\left[\Delta\boldsymbol{Y}_{Ly}^T(k)\quad \Delta\boldsymbol{U}_{Lu}^T(k)\right]^T$;

$\Delta\boldsymbol{Y}_{Ly}(k)=[\Delta\boldsymbol{y}^T(k),\cdots,\Delta\boldsymbol{y}^T(k-L_y+1)]^T$;

$\Delta\boldsymbol{U}_{Lu}(k)=[\Delta\boldsymbol{u}^T(k),\cdots,\Delta\boldsymbol{u}^T(k-L_u+1)]^T$;

The positive integers $L_y$ ($0\leq L_y$) and $L_u$ ($1\leq L_u$) are called pseudo orders.

*Proof*: The proof is similar to Theorem 1 and we omit it.

### *B. Design of Model-Free Adaptive Control for MIMO systems*

We can rewrite (30) into (31).

$$\boldsymbol{y}(k+1)=\boldsymbol{y}(k)+\boldsymbol{\phi}_L^T(k)\Delta\boldsymbol{H}(k)+\Delta\boldsymbol{w}(k+1) \tag{31}$$

The object is to design a controller that optimizes the cost function:

$$J=\left[\boldsymbol{y}^*(k+1)-\boldsymbol{y}(k+1)\right]^T\left[\boldsymbol{y}^*(k+1)-\boldsymbol{y}(k+1)\right]+\Delta\boldsymbol{u}^T(k)\boldsymbol{\lambda}\Delta\boldsymbol{u}(k) \tag{32}$$

where $\boldsymbol{\lambda}=\mathrm{diag}(\lambda_1,\cdots,\lambda_{Mu})$ is the weighted diagonal matrix and we assume $\lambda_i$($i=1,\cdots,M_u$) are equal to $\lambda$ according to [2]; $\boldsymbol{y}^*(k+1)=\left[y_1^*(k+1),\cdots,y_{My}^*(k+1)\right]^T$ is the desired trajectory vector.

Substituting (31) into (32) and solving the optimization condition $\partial J/\partial\Delta\boldsymbol{u}(k)=0$ yield

$$\begin{aligned}&[\boldsymbol{\Phi}_{Ly+1}^T(k)\boldsymbol{\Phi}_{Ly+1}(k)+\boldsymbol{\lambda}]\Delta\boldsymbol{u}(k)=\boldsymbol{\Phi}_{Ly+1}^T(k)[(\boldsymbol{y}^*(k+1)-\boldsymbol{y}(k))\\&-\sum_{i=1}^{Ly}\boldsymbol{\Phi}_i(k)\Delta\boldsymbol{y}(k-i+1)-\sum_{i=Ly+2}^{Ly+Lu}\boldsymbol{\Phi}_i(k)\Delta\boldsymbol{u}(k-i+1)-\Delta\boldsymbol{w}(k+1)]\end{aligned} \tag{33}$$

Since the disturbance $\boldsymbol{w}(k)$ may not be acquired directly, we replace it by $\hat{\boldsymbol{w}}(k)$ which represents the estimation of the disturbance. Then we rewrite (33) into (34).

$$\begin{aligned}&\Delta\boldsymbol{u}(k)=[\boldsymbol{\Phi}_{Ly+1}^T(k)\boldsymbol{\Phi}_{Ly+1}(k)+\boldsymbol{\lambda}]^{-1}\boldsymbol{\Phi}_{Ly+1}^T(k)[(\boldsymbol{y}^*(k+1)-\boldsymbol{y}(k))\\&-\sum_{i=1}^{Ly}\boldsymbol{\Phi}_i(k)\Delta\boldsymbol{y}(k-i+1)-\sum_{i=Ly+2}^{Ly+Lu}\boldsymbol{\Phi}_i(k)\Delta\boldsymbol{u}(k-i+1)-\Delta\hat{\boldsymbol{w}}(k+1)]\end{aligned} \tag{34}$$

We define

$$\boldsymbol{\phi}_{Ly}(z^{-1})=\boldsymbol{\Phi}_1(k)+\cdots+\boldsymbol{\Phi}_{Ly}(k)z^{-Ly+1} \tag{35}$$

$$\boldsymbol{\phi}_{Lu}(z^{-1})=\boldsymbol{\Phi}_{Ly+1}(k)+\cdots+\boldsymbol{\Phi}_{Ly+Lu}(k)z^{-Lu+1} \tag{36}$$

Then (30) is rewritten as

$$\Delta\boldsymbol{y}(k+1)=\boldsymbol{\phi}_{Ly}(z^{-1})\Delta\boldsymbol{y}(k)+\boldsymbol{\phi}_{Lu}(z^{-1})\Delta\boldsymbol{u}(k)+\Delta\boldsymbol{w}(k+1) \tag{37}$$

From (34)-(37), we can have the closed-loop system equation (38) at the time $k$:

$$\begin{aligned}&\left[\Delta\lambda\left[\boldsymbol{I}-z^{-1}\boldsymbol{\phi}_{Ly}(z^{-1})\right]+\boldsymbol{\phi}_{Lu}(z^{-1})\boldsymbol{\Phi}_{Ly+1}^T(k)\right]\boldsymbol{y}(k)\\&=\boldsymbol{\phi}_{Lu}(z^{-1})\boldsymbol{\Phi}_{Ly+1}^T(k)\boldsymbol{y}^*(k)+[\boldsymbol{\lambda}+\boldsymbol{\phi}_{Lu}(z^{-1})\boldsymbol{\Phi}_{Ly+1}^T(k)]\Delta\boldsymbol{w}(k)\\&-\boldsymbol{\phi}_{Lu}(z^{-1})\boldsymbol{\Phi}_{Ly+1}^T(k)\Delta\hat{\boldsymbol{w}}(k)\end{aligned} \tag{38}$$

Assume $\mathrm{rank}\left[\boldsymbol{\Phi}_{Ly+1}(k)\right]=M_y$ ($M_u\geq M_y$), we may obtain the inequality (39) by tuning $\boldsymbol{\lambda}$.

$$\boldsymbol{T}=\Delta\lambda\left[\boldsymbol{I}-z^{-1}\boldsymbol{\phi}_{Ly}(z^{-1})\right]+\boldsymbol{\phi}_{Lu}(z^{-1})\boldsymbol{\Phi}_{Ly+1}^T(k)\neq\boldsymbol{0},\quad |z|>1 \tag{39}$$

(39) determines the poles of the system.

If $\Delta\hat{\boldsymbol{w}}(k+1)=\Delta\boldsymbol{w}(k+1)$, the disturbance-to-output transfer function will be

$$\boldsymbol{G}(z^{-1})=\frac{\boldsymbol{\lambda}(1-z^{-1})}{\boldsymbol{\lambda}(1-z^{-1})\left[\boldsymbol{I}-z^{-1}\boldsymbol{\phi}_{Ly}(z^{-1})\right]+\boldsymbol{\phi}_{Lu}(z^{-1})\boldsymbol{\Phi}_{Ly+1}^T(k)} \tag{40}$$

Additionally, when $\boldsymbol{\lambda}=\boldsymbol{0}$, the closed-loop system equation will be

$$\begin{aligned}&\left[\Delta\lambda\left[\boldsymbol{I}-z^{-1}\boldsymbol{\phi}_{Ly}(z^{-1})\right]+\boldsymbol{\phi}_{Lu}(z^{-1})\boldsymbol{\Phi}_{Ly+1}^T(k)\right]\boldsymbol{y}(k)\\&=\boldsymbol{\phi}_{Lu}(z^{-1})\boldsymbol{\Phi}_{Ly+1}^T(k)\boldsymbol{y}^*(k)\end{aligned} \tag{41}$$

which indicates that the influence of disturbance $\boldsymbol{w}(k)$ will be theoretically removed when (39) is satisfied.

On the other hand, if $\boldsymbol{w}(k+1)$ is unknown, we normally let $\Delta\hat{\boldsymbol{w}}(k+1)=\boldsymbol{0}$ in the controller design process, and the closed-loop system equations at the time of $k$ will be

$$\begin{aligned}&\left[\Delta\lambda\left[\boldsymbol{I}-z^{-1}\boldsymbol{\phi}_{Ly}(z^{-1})\right]+\boldsymbol{\phi}_{Lu}(z^{-1})\boldsymbol{\Phi}_{Ly+1}^T(k)\right]\boldsymbol{y}(k)\\&=\boldsymbol{\phi}_{Lu}(z^{-1})\boldsymbol{\Phi}_{Ly+1}^T(k)\boldsymbol{y}^*(k)+[\boldsymbol{\lambda}+\boldsymbol{\phi}_{Lu}(z^{-1})\boldsymbol{\Phi}_{Ly+1}^T(k)]\Delta\boldsymbol{w}(k)\end{aligned} \tag{42}$$

As $\boldsymbol{\lambda}=\boldsymbol{0}$, the transfer function for the disturbance is $(1-z^{-1})\boldsymbol{I}$.

Similarly, the disturbance can be estimated by

$$\hat{\boldsymbol{w}}(k+1)=\hat{\boldsymbol{w}}(k)-\boldsymbol{L}\left[\hat{\boldsymbol{w}}(k)-\boldsymbol{w}(k)\right] \tag{43}$$

$$\boldsymbol{w}(k)=\boldsymbol{f}(\boldsymbol{\varphi}(k))-\boldsymbol{y}(k) \tag{44}$$

where $\boldsymbol{L}=diag(l_1,\cdots,l_{My})$, and the estimator will be stable when $l_i\in[0, 2]$, ($i=1,\cdots, M_y$). Then we have the relationship between the disturbance and its estimation at the time $k$:

$$\hat{\boldsymbol{w}}(k+1)=[\boldsymbol{I}+(\boldsymbol{L}-\boldsymbol{I})z^{-1}]^{-1}\boldsymbol{L}\boldsymbol{w}(k) \tag{45}$$

### *C. Simulations:*

Example 3: We consider the following MIMO nonlinear system:

$$\begin{aligned} y_1(k+1) &= -0.7y_1^3(k) + y_2^2(k) + u_1(k) + 0.4u_2(k) \\ &\quad + 0.1u_1^2(k-1) + 0.2u_2^4(k-1) + w_1(k+1) \\ y_2(k+1) &= -0.9y_1^2(k) + 0.8y_2^3(k) + 0.5u_1(k) + 1.1u_2(k) \\ &\quad - 0.1u_1^3(k-1) + 0.1u_2^2(k-1) + w_2(k+1) \end{aligned} \tag{46}$$

where the known disturbance vector is

$$\boldsymbol{w}(k+1) = \begin{bmatrix} w_1(k+1) \\ w_2(k+1) \end{bmatrix} = \begin{bmatrix} \sin(k/10) \\ \cos(k/30) \end{bmatrix} \tag{47}$$

The desired trajectories are

$$\begin{aligned} & y_1^*(k+1) = 0.3\sin(k/40) - 0.1\cos(k/5) \qquad 1 \le k \le 400 \\ & y_2^*(k+1) = 0.2\sin(k/10) - 0.3\cos(k/30) \qquad 1 \le k \le 400 \\ & y_1^*(k+1) = -y_2^*(k+1) = 0.1 \times (-1)^{round(k/50)} \quad 401 \le k \le 800 \end{aligned} \tag{48}$$

The initial values are $\boldsymbol{y}(1)=\boldsymbol{y}(2)=\boldsymbol{y}(3)=[0,0]^T$. According to [4]-[6], the controller structure should be applied with $L_y=n_y+1=1$, $L_u=n_u+1=2$. The elements of PJM are calculated by

$$\hat{\boldsymbol{\Phi}}_1(k) = \begin{bmatrix} \hat{\phi}_{11}(k) & \hat{\phi}_{12}(k) \\ \hat{\phi}_{21}(k) & \hat{\phi}_{22}(k) \end{bmatrix} \quad ; \quad \hat{\boldsymbol{\Phi}}_3(k) = \begin{bmatrix} \hat{\phi}_{15}(k) & \hat{\phi}_{16}(k) \\ \hat{\phi}_{25}(k) & \hat{\phi}_{26}(k) \end{bmatrix} \quad ;$$

$$\hat{\boldsymbol{\Phi}}_2(k) = \begin{bmatrix} \hat{\phi}_{13}(k) & \hat{\phi}_{14}(k) \\ \hat{\phi}_{23}(k) & \hat{\phi}_{24}(k) \end{bmatrix} = \begin{bmatrix} 1 & 0.4 \\ 0.5 & 1.1 \end{bmatrix};$$

$$\hat{\phi}_{11}(k) = -0.7(3y_1^2(k-1) + 3y_1(k-1)\Delta y_1(k-1) + \Delta y_1^2(k-1)) \;;$$

$$\hat{\phi}_{12}(k) = 2y_2(k-1) + \Delta y_2(k-1) \;;$$

$$\hat{\phi}_{21}(k) = 0.9(2y_1(k-1) + \Delta y_1(k-1)) \;;$$

$$\hat{\phi}_{22}(k) = 0.8(3y_2^2(k-1) + 3y_2(k-1)\Delta y_2(k-1) + \Delta y_2^2(k-1)) \;;$$

$$\hat{\phi}_{15}(k) = 0.1(2u_1(k-2) + \Delta u_1(k-2)) \;;$$

$$\begin{aligned} \hat{\phi}_{16}(k) = 0.2(&4u_2^3(k-2) + 6u_2^2(k-2)\Delta u_2((k-2) \\ &+ 4u_2(k-2)\Delta u_2^2((k-2) + \Delta u_2^3((k-2)) \end{aligned} \;;$$

$$\hat{\phi}_{25}(k) = -0.1(3u_1^2(k-2) + 3u_1(k-2)\Delta u_1(k-2) + \Delta u_1^2(k-2)) \;;$$

$$\hat{\phi}_{26}(k) = 0.1(2u_2(k-2) + \Delta u_2(k-2)) \;;$$

We made the comparisons between the controller (33) with $\boldsymbol{\lambda}=\boldsymbol{0}$, $\boldsymbol{\lambda}=0.5\boldsymbol{I}$ and $\boldsymbol{\lambda}=1.5\boldsymbol{I}$. Fig. 7 and Fig. 8 show the tracking performance. Fig. 9 shows the control inputs. Fig. 10 shows the elements in the calculated PJM.

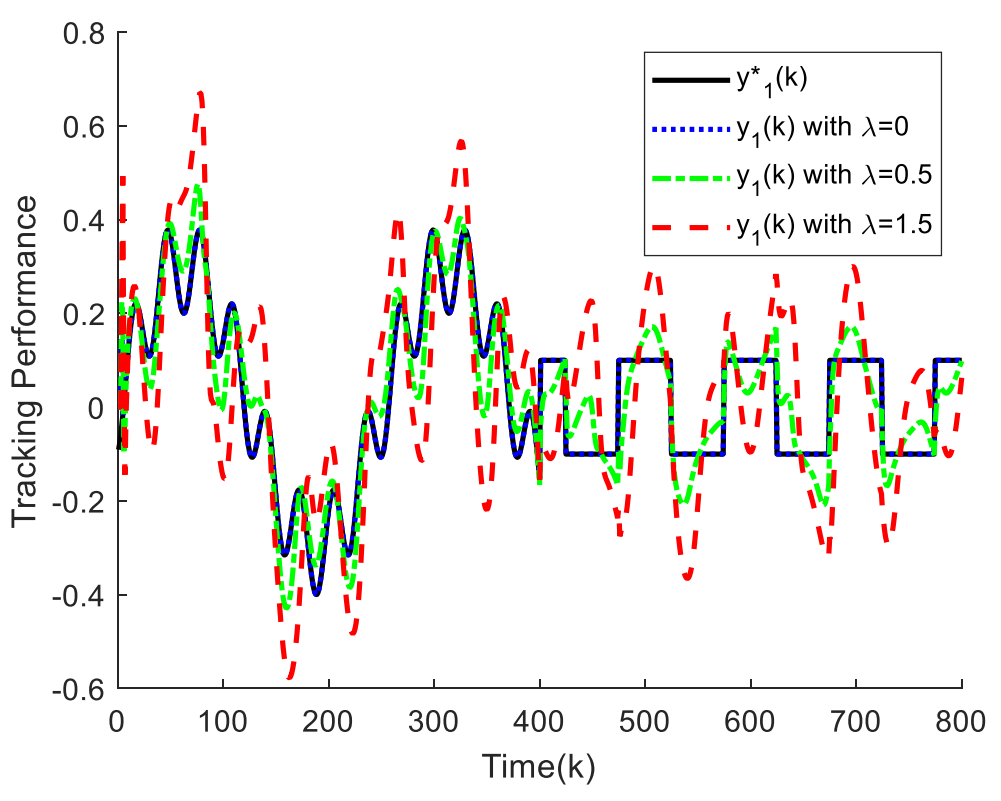

Fig. 7 Tracking performance of $y_1$

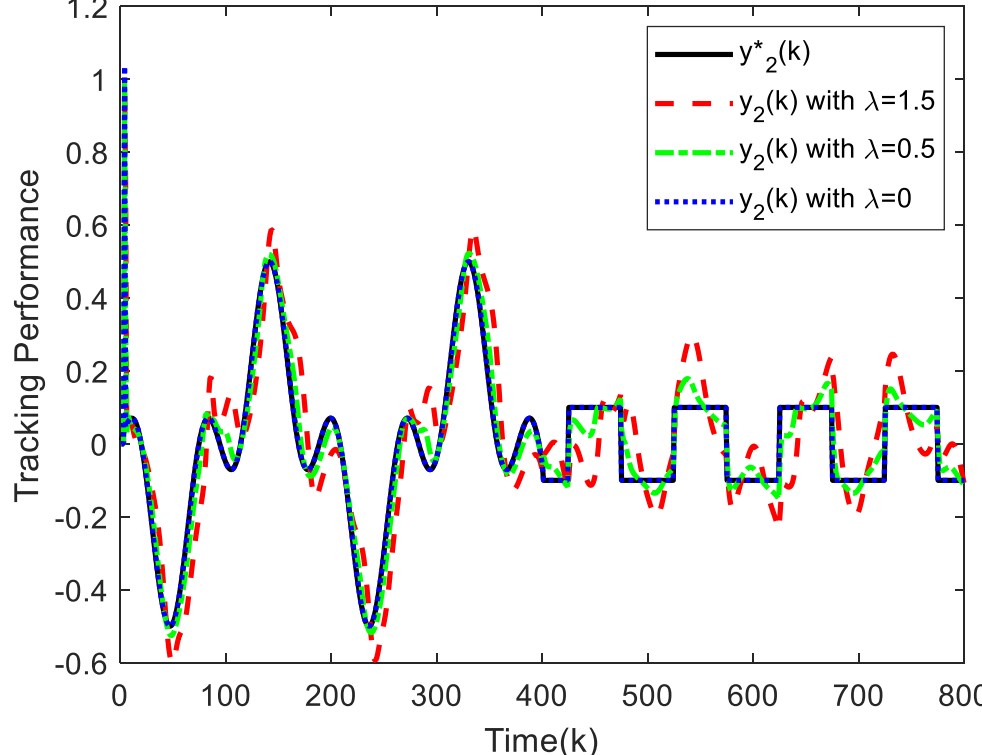

Fig. 8 Tracking performance of *y*2

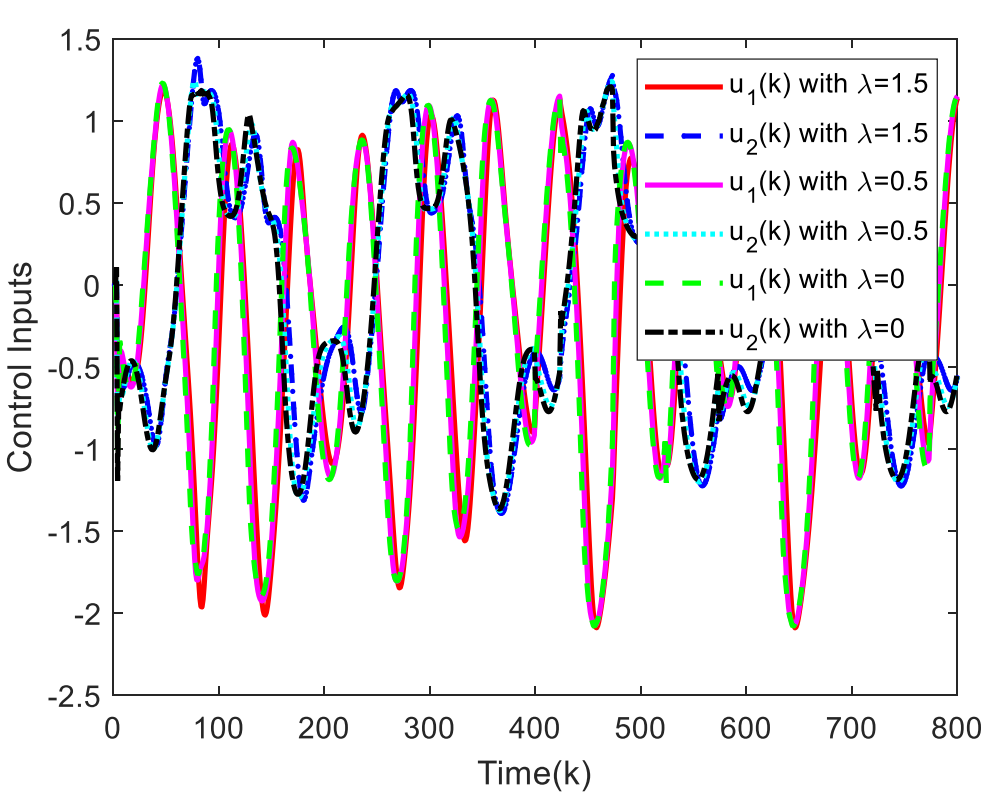

Fig. 9 Control inputs

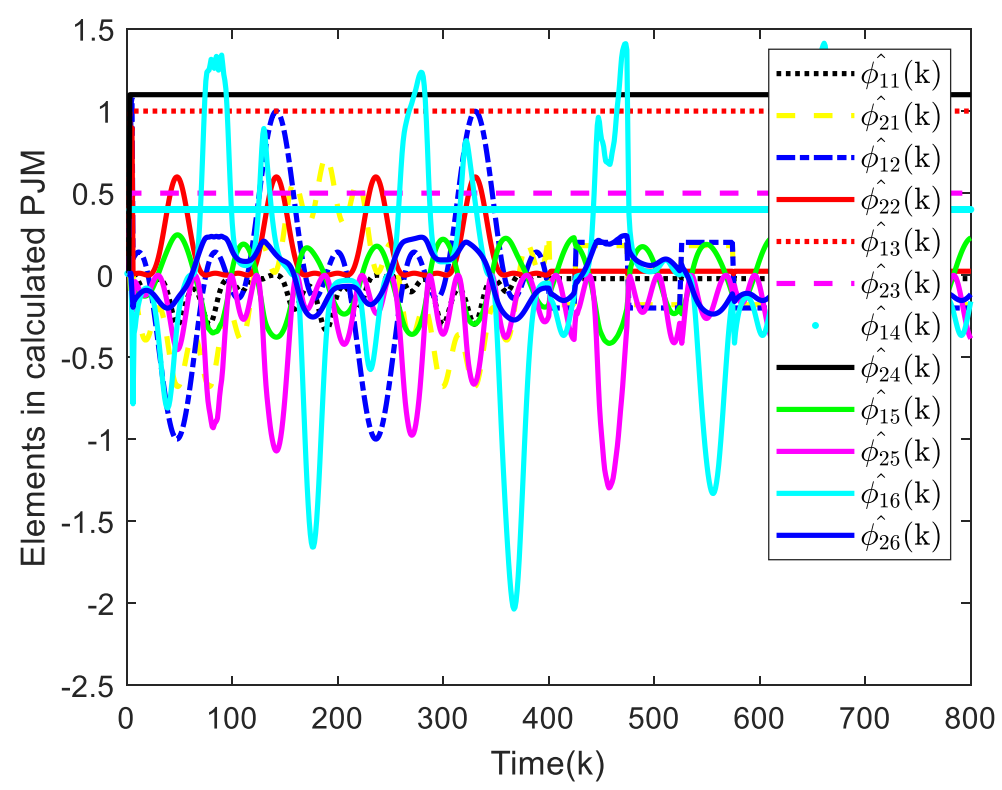

Fig. 10 Elements in calculated PJM

In this example, Fig. 7 and Fig. 8 obviously show a phenomenon that the disturbance rejection effect will decrease when $\lambda$ raises. Further, the influence of disturbance is removed only when the system is controlled by (33) with $\boldsymbol{\lambda}=\boldsymbol{0}$.

*Example 4*: In this example, we will show that Theorem 4 in [14] is irrational. To make clear its essence, we study only one separate agent and let $\boldsymbol{y}(k+1)=\boldsymbol{r}(k)+\Delta\boldsymbol{r}(k+1)$, $\sum_{j\in N_i} a_{ij}=0$, $d_i=1$ and $\Delta\boldsymbol{w}(k+1)=\boldsymbol{\Phi}_{i,3}(k)\Delta\boldsymbol{w}_i(k)$ in [14], then the controller (9) in [14] becomes the controller (33) in this example.

We consider the following MIMO linear system:

$$\begin{aligned}
&\boldsymbol{y}(k+1) = \boldsymbol{f}(\boldsymbol{y}(k), \boldsymbol{y}(k-1), \boldsymbol{u}(k), \boldsymbol{w}(k+1)) \\
&= \boldsymbol{\Phi}_1(k)\Delta\boldsymbol{y}(k) + \boldsymbol{\Phi}_2(k)\Delta\boldsymbol{y}(k-1) + \boldsymbol{\Phi}_3(k)\Delta\boldsymbol{u}(k) + \boldsymbol{w}(k+1) \\
&= \begin{bmatrix} -1 & 2 \\ -1 & -1.4 \end{bmatrix}\begin{bmatrix} y_1(k) \\ y_2(k) \end{bmatrix} + \begin{bmatrix} 0.6 & 6 \\ 0.6 & -3 \end{bmatrix}\begin{bmatrix} y_1(k-1) \\ y_2(k-1) \end{bmatrix} + \begin{bmatrix} 1.3 & 1 \\ 1 & 0 \end{bmatrix}\begin{bmatrix} u_1(k) \\ u_2(k) \end{bmatrix} \\
&\quad + \boldsymbol{w}(k+1)
\end{aligned} \tag{49}$$

where $\boldsymbol{w}(k+1) = \begin{bmatrix} 20\sin(k/20) + 40\cos(k/40) + 9e^{k/100} \\ 20\cos(k/30) + 40\cos(k/50) + 20e^{k/150} \end{bmatrix}$.

The desired trajectories are

$$y_1^*(k+1) = y_2^*(k+1) = 3\times(-1)^{round(k/50)} \tag{50}$$

The initial values are $\boldsymbol{y}(1)=\boldsymbol{y}(3)=[0,0]^T$, $\boldsymbol{y}(2)=[1,1]^T$. According to [4]-[6] and [15], the controller structure should be applied with $L_y=n_y+1=2$, $L_u=n_u+1=1$, and the controller coefficients are set in accordance with the actual system model (49) to comprehend its nature more exactly. We made the comparisons between the controller (33) with $\boldsymbol{\lambda}=\boldsymbol{0}$, $\boldsymbol{\lambda}=0.02\boldsymbol{I}$ and the original MFAC controller (51) with $\boldsymbol{\lambda}=\boldsymbol{0}$.

$$\begin{aligned}
\Delta\boldsymbol{u}(k) = {} & [\boldsymbol{\Phi}_3^T(k)\boldsymbol{\Phi}_3(k) + \boldsymbol{\lambda}]^{-1}\boldsymbol{\Phi}_2^T(k)[(\boldsymbol{y}^*(k+1) - \boldsymbol{y}(k)) \\
& - \sum_{i=1}^{2}\boldsymbol{\Phi}_i(k)\Delta\boldsymbol{y}(k-i+1)]
\end{aligned} \tag{51}$$

Fig. 11 and Fig. 12 show the tracking performance. Fig. 13 shows the control inputs.

From Fig. 11 and Fig. 12, we can see that the influence of disturbance is removed only when the system is controlled by (33) with $\lambda=0$. When $\lambda$ raises to 0.02, the control effect is poor.

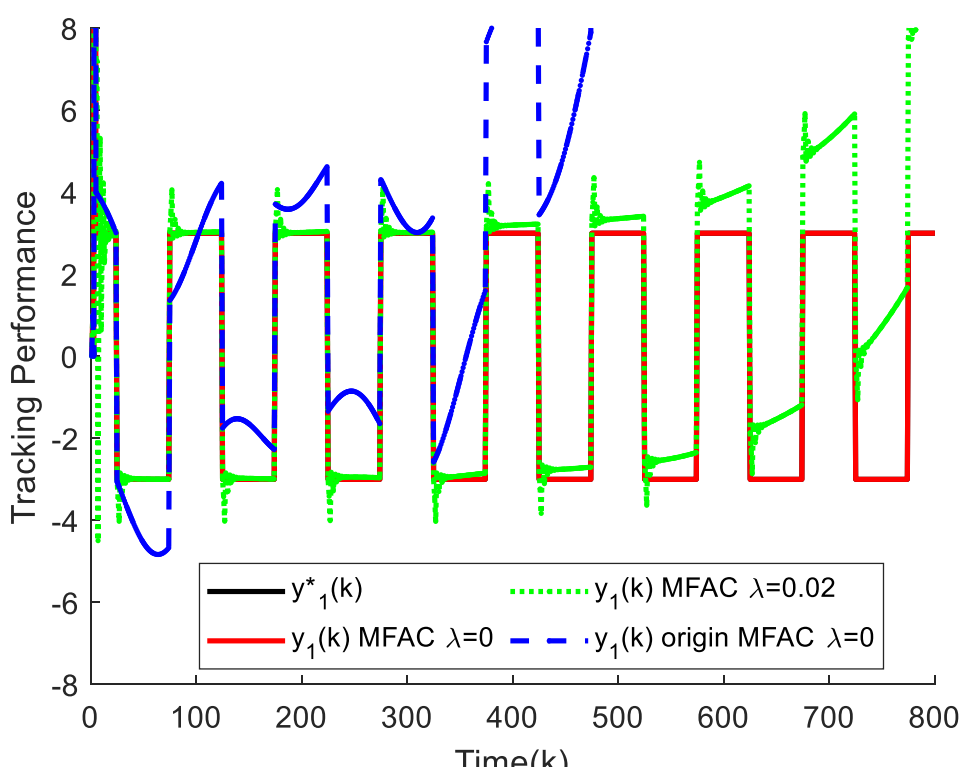

Fig. 11 Tracking performance of $y_1$

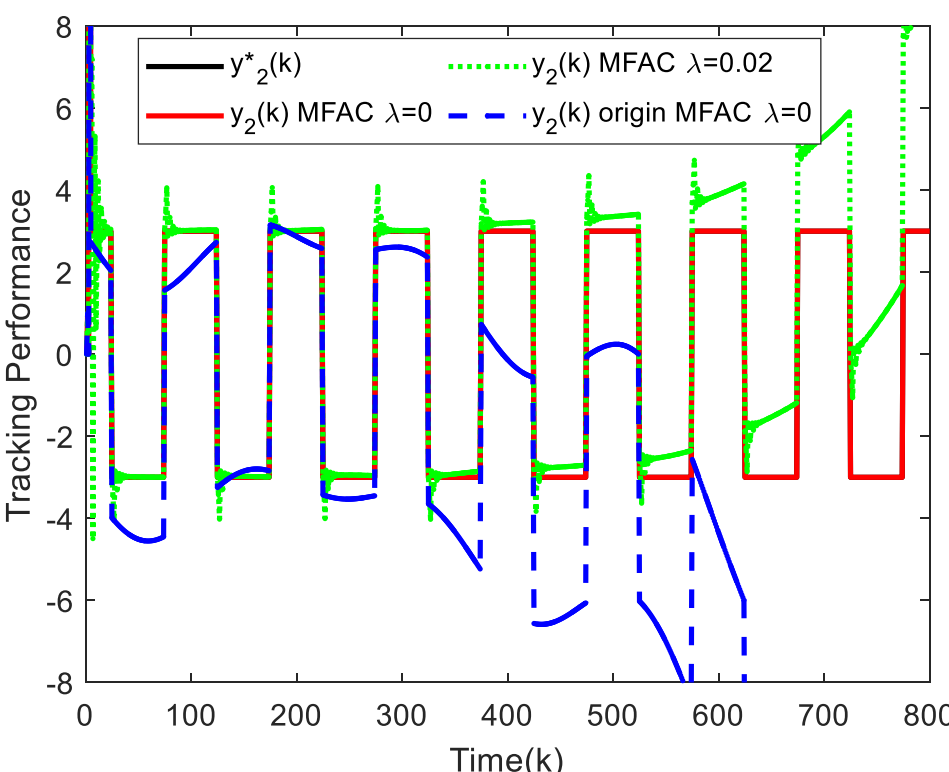

Fig. 12 Tracking performance of $y2$

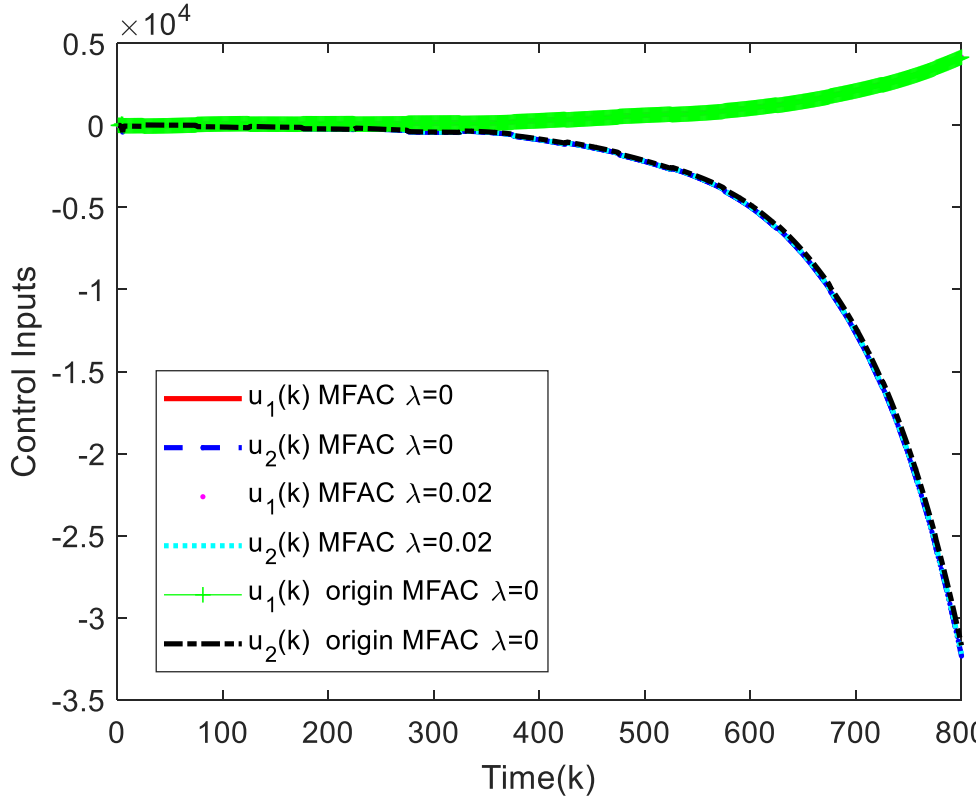

Fig. 13 Control inputs

If we choose controller (33) with $\lambda\geq0.1$, the system output will be divergent. To figure out the reason behind the phenomenon, we consider system (49) with $\boldsymbol{w}(k)=[0,0]^T$ and choose the controller (51) with $\lambda\geq0.1$, the system outputs are divergent. Fig. 14 shows one system output $y_1(k)$ as $\lambda=0.1$. Therefore, the conclusion that "tracking errors are bounded when $\lambda>\lambda_{\min}$" in [14] is not correct. On the contrary, the influence of disturbance will be removed and the tracking error will converge very fast when $\lambda=0$ in this case.

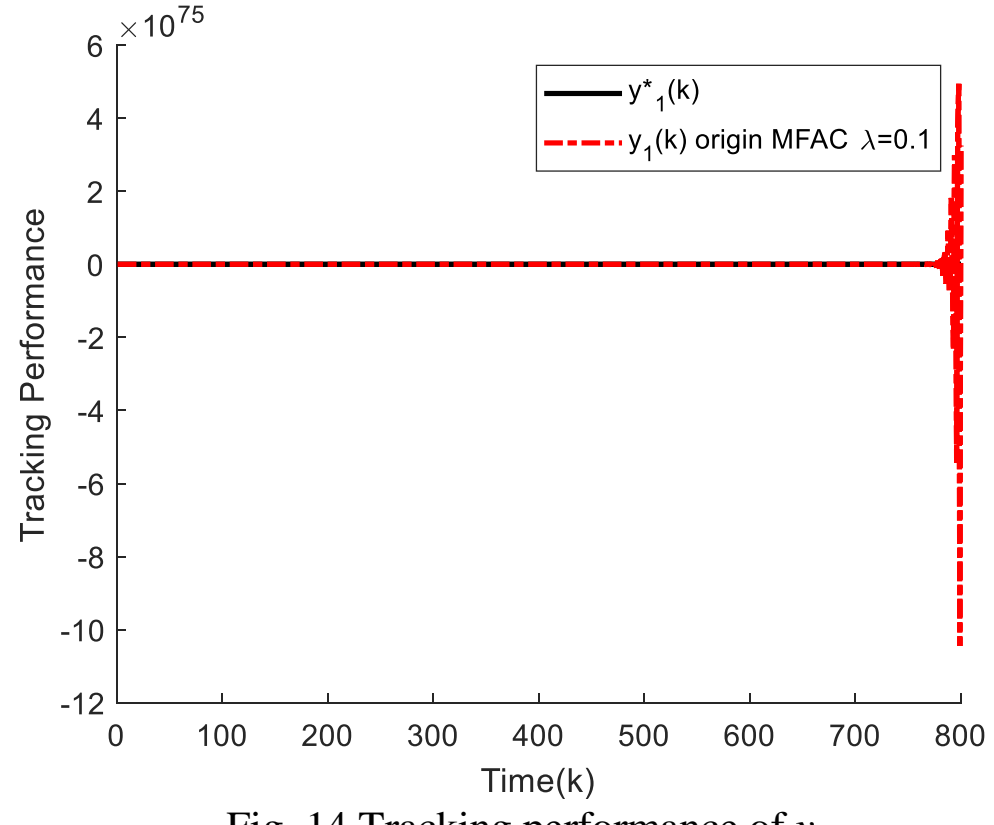

Fig. 14 Tracking performance of $y_1$

## IV. CONCLUSION

In this paper, we modify the EDLM with disturbance and prove it according to the definition of differentiability. Based on the modified EDLM, we redesign MFAC compensated with disturbances and analyze the discrete-time nonlinear system by the transient closed-loop system equation at each time. This is all possible because some nonlinear system functions can be accurately described by the EDLM compensated with disturbance according to the Taylor series or the definition of differentiability. Finally, several examples are provided to validate our viewpoints. At the end of this conclusion, it is evident that renaming the MFAC compensated with disturbance is necessary. 'Incremental one-step-ahead control compensated with disturbance' emerges as a potential alternative, but the ideal term is still open for discussion. I believe that a more appropriate name will be adopted in the future.

## APPENDIX

Proof of *Theorem 1*

*Proof*: Case 1: $1\leq L_y\leq n_y$ and $1\leq L_u\leq n_u$

From (1), we have

$$\begin{aligned}
&\Delta y(k+1)= \\
&f(y(k),\cdots,y(k-L_y+1),y(k-L_y)\cdots,y(k-n_y),u(k), \\
&\quad \cdots,u(k-L_u+1),u(k-L_u),\cdots,u(k-n_u))+w(k+1) \\
&-f(y(k-1),\cdots,y(k-L_y),y(k-L_y),\cdots,y(k-n_y),u(k-1), \\
&\quad \cdots,u(k-L_u),u(k-L_u),\cdots,u(k-n_u)) \\
&+f(y(k-1),\cdots,y(k-L_y),y(k-L_y),\cdots,y(k-n_y),u(k-1), \\
&\quad \cdots,u(k-L_u),u(k-L_u),\cdots,u(k-n_u)) \\
&-f(y(k-1),\cdots,y(k-L_y),y(k-L_y-1),\cdots,y(k-n_y-1), \\
&\quad u(k-1),\cdots,u(k-L_u),u(k-L_u-1),\cdots,u(k-n_u-1))-w(k)
\end{aligned} \tag{52}$$

According to the definition of differentiability in [16]-[17], (52) becomes

$$\begin{aligned}
\Delta y(k+1)=&\frac{\partial f(\boldsymbol{\varphi}(k-1))}{\partial y(k-1)}\Delta y(k)+\cdots+\frac{\partial f(\boldsymbol{\varphi}(k-1))}{\partial y(k-L_y)}\Delta y(k-L_y+1) \\
&+\frac{\partial f(\boldsymbol{\varphi}(k-1))}{\partial u(k-1)}\Delta u(k)+\cdots+\frac{\partial f(\boldsymbol{\varphi}(k-1))}{\partial u(k-L_u)}\Delta u(k-L_u+1) \\
&+\varepsilon_1(k)\Delta y(k)+\cdots+\varepsilon_{Ly}(k)\Delta y(k-L_y+1)+\varepsilon_{Ly+1}(k)\Delta u(k) \\
&+\cdots+\varepsilon_{Ly+Lu}(k)\Delta u(k-L_u+1)+\psi(k)+\Delta w(k+1)
\end{aligned} \tag{53}$$

where

$$\begin{aligned}
\psi(k)\triangleq\ & f(y(k-1),\cdots,y(k-L_y),y(k-L_y),\cdots,y(k-n_y), \\
&u(k-1),\cdots,u(k-L_u),u(k-L_u),\cdots,u(k-n_u)) \\
&-f(y(k-1),\cdots,y(k-L_y),y(k-L_y-1),\cdots,y(k-n_y-1), \\
&u(k-1),\cdots,u(k-L_u),u(k-L_u-1),\cdots,u(k-n_u-1))
\end{aligned} \tag{54}$$

$\frac{\partial f(\boldsymbol{\varphi}(k-1))}{\partial y(k-i-1)}$, $0\le i\le L_y-1$ and $\frac{\partial f(\boldsymbol{\varphi}(k-1))}{\partial u(k-j-1)}$, $0\le j\le L_u-1$ denote the partial derivative values of $f(\boldsymbol{\varphi}(k-1))$ with respect to the ($i$+1)-th variable and the ($n_y$+2+$j$)-th variable, respectively. And $\varepsilon_1(k),\cdots,\varepsilon_{Ly+Lu}(k)$ are functions that depend only on $\Delta y(k),\cdots,\Delta y(k-L_y+1),\Delta u(k),\cdots,\Delta u(k-L_u+1)$, with $(\varepsilon_1(k),\cdots,\varepsilon_{Ly+Lu}(k))\to(0,\cdots,0)$ when $(\Delta y(k),\cdots,\Delta y(k-L_y+1),\Delta u(k),\cdots,\Delta u(k-L_u+1))\to(0,\cdots,0)$. This also implies that $(\varepsilon_1(k),\cdots,\varepsilon_{Ly+Lu}(k))$ can be regarded as $(0,\cdots,0)$ when the control period of the system is sufficiently small.

We consider the following equation with the vector $\boldsymbol{\eta}(k)$ for each time $k$:

$$\psi(k)=\boldsymbol{\eta}^T(k)\Delta\boldsymbol{H}(k) \tag{55}$$

Owing to $\|\Delta\boldsymbol{H}(k)\|\neq 0$, (55) must have at least one solution $\boldsymbol{\eta}^*(k)$. Let

$$\begin{aligned}
\boldsymbol{\phi}_L(k)=\boldsymbol{\eta}^*(k)+[&\frac{\partial f(\boldsymbol{\varphi}(k-1))}{\partial y(k-1)}+\varepsilon_1(k),\cdots,\frac{\partial f(\boldsymbol{\varphi}(k-1))}{\partial y(k-L_y)}+\varepsilon_{Ly}(k), \\
&\frac{\partial f(\boldsymbol{\varphi}(k-1))}{\partial u(k-1)}+\varepsilon_{Ly+1}(k),\cdots,\frac{\partial f(\boldsymbol{\varphi}(k-1))}{\partial u(k-L_u)}+\varepsilon_{Ly+Lu}(k)]^T
\end{aligned} \tag{56}$$

(53) can be described as follows:

$$\Delta y(k+1)=\boldsymbol{\phi}_L^T(k)\Delta\boldsymbol{H}(k)+\Delta w(k+1) \tag{57}$$

Case 2: $L_y$=$n_y$+1 and $L_u$=$n_u$ +1

According to the definition of differentiability in [16]-[17], (1) becomes

$$\begin{aligned}
\Delta y(k+1)=&\frac{\partial f(\boldsymbol{\varphi}(k-1))}{\partial y(k-1)}\Delta y(k)+\cdots+\frac{\partial f(\boldsymbol{\varphi}(k-1))}{\partial y(k-n_y-1)}\Delta y(k-n_y) \\
&+\frac{\partial f(\boldsymbol{\varphi}(k-1))}{\partial u(k-1)}\Delta u(k)+\cdots+\frac{\partial f(\boldsymbol{\varphi}(k))}{\partial u(k-n_u-1)}\Delta u(k-n_u) \\
&+\gamma(k)+\Delta w(k+1)
\end{aligned} \tag{58}$$

where

$$\begin{aligned}
\gamma(k)=\ &\varepsilon_1(k)\Delta y(k)+\cdots+\varepsilon_{Ly}(k)\Delta y(k-n_y) \\
&+\varepsilon_{Ly+1}(k)\Delta u(k)+\cdots+\varepsilon_{Ly+Lu}(k)\Delta u(k-n_u)
\end{aligned} \tag{59}$$

We let

$$\begin{aligned}
\boldsymbol{\phi}_L(k)=[&\frac{\partial f(\boldsymbol{\varphi}(k-1))}{\partial y(k-1)}+\varepsilon_1(k),\cdots,\frac{\partial f(\boldsymbol{\varphi}(k-1))}{\partial y(k-n_y-1)}+\varepsilon_{Ly}(k), \\
&\frac{\partial f(\boldsymbol{\varphi}(k-1))}{\partial u(k-1)}+\varepsilon_{Ly+1}(k),\cdots,\frac{\partial f(\boldsymbol{\varphi}(k-1))}{\partial u(k-n_u-1)}+\varepsilon_{Ly+Lu}(k)]^T
\end{aligned} \tag{60}$$

to rewrite (58) as (57), with $(\varepsilon_1(k),\cdots,\varepsilon_{Ly+Lu}(k))\to(0,\cdots,0)$ in nonlinear systems, when $(\Delta y(k),\cdots,\Delta y(k-n_y),\Delta u(k),\cdots,\Delta u(k-n_u))\to(0,\cdots,0)$. As to linear systems, we will always have $\boldsymbol{\phi}_L(k)=[\frac{\partial f(\boldsymbol{\varphi}(k-1))}{\partial y(k-1)},\cdots,\frac{\partial f(\boldsymbol{\varphi}(k-1))}{\partial y(k-n_y-1)},\frac{\partial f(\boldsymbol{\varphi}(k-1))}{\partial u(k-1)},\cdots,\frac{\partial f(\boldsymbol{\varphi}(k-1))}{\partial u(k-n_u-1)}]^T$ no matter what $(\Delta y(k),\cdots,\Delta y(k-n_y),\Delta u(k),\cdots,\Delta u(k-n_u))$ is.

Additionally, if the function $f(\cdots)$ has derivatives of all orders on any operating points, we can obtain (61) in accordance with the Taylor series:

$$\begin{aligned}
\Delta y(k+1)=[&\Delta y(k)\frac{\partial}{\partial y(k-1)}+\cdots+\Delta y(k-n_y)\frac{\partial}{\partial y(k-n_y-1)} \\
&+\Delta u(k)\frac{\partial}{\partial u(k-1)}+\cdots+\Delta u(k-n_u)\frac{\partial}{\partial u(k-n_u-1)}]f(\boldsymbol{\varphi}(k-1)) \\
&+\cdots+\frac{1}{n!}[\Delta y(k)\frac{\partial}{\partial y(k-1)}+\cdots+\Delta y(k-n_y)\frac{\partial}{\partial y(k-n_y-1)} \\
&+\Delta u(k)\frac{\partial}{\partial u(k-1)}+\cdots+\Delta u(k-n_u)\frac{\partial}{\partial u(k-n_u-1)}]^n f(\boldsymbol{\varphi}(k-1)) \\
&+\cdots
\end{aligned} \tag{61}$$

and obtain a group of solutions (62), (63) for (59) from (61).

$$\begin{aligned}
\varepsilon_{i+1}(k)=&\frac{1}{2!}\frac{\partial^2 f(\boldsymbol{\varphi}(k-1))}{\partial y^2(k-i-1)}\Delta y(k-i)+\frac{1}{3!}\frac{\partial^3 f(\boldsymbol{\varphi}(k-1))}{\partial y^3(k-i-1)}\Delta y^2(k-i) \\
&+\frac{1}{4!}\frac{\partial^4 f(\boldsymbol{\varphi}(k-1))}{\partial y^4(k-i-1)}\Delta y^3(k-i)+\cdots
\end{aligned} \tag{62}$$

$$\varepsilon_{Ly+1+j}(k)=\frac{1}{2!}\frac{\partial^2 f(\boldsymbol{\varphi}(k-1))}{\partial u^2(k-j-1)}\Delta u(k-j)+\frac{1}{3!}\frac{\partial^3 f(\boldsymbol{\varphi}(k-1))}{\partial u^3(k-j-1)} \bullet\Delta u^2(k-j)+\frac{1}{4!}\frac{\partial^4 f(\boldsymbol{\varphi}(k-1))}{\partial u^4(k-j-1)}\Delta u^3(k-j)+\cdots \tag{63}$$

, $i=0,\cdots,n_y$ and $j=0,\cdots,n_u$.

Case 3: $L_y>n_y+1$ and $L_u>n_u+1$

According to the definition of differentiability in [16]-[17], (1) becomes

$$\begin{aligned}\Delta y(k+1)=&\frac{\partial f(\boldsymbol{\varphi}(k-1))}{\partial y(k-1)}\Delta y(k)+\cdots+\frac{\partial f(\boldsymbol{\varphi}(k-1))}{\partial y(k-n_y-1)}\Delta y(k-n_y)\\&+\frac{\partial f(\boldsymbol{\varphi}(k-1))}{\partial u(k-1)}\Delta u(k)+\cdots+\frac{\partial f(\boldsymbol{\varphi}(k-1))}{\partial u(k-n_u-1)}\Delta u(k-n_u)\\&+\varepsilon_1(k)\Delta y(k)+\cdots+\varepsilon_{n_y+1}(k)\Delta y(k-n_y)+\varepsilon_{Ly+1}(k)\Delta u(k)\\&+\cdots+\varepsilon_{Ly+n_u+1}(k)\Delta u(k-n_u)+\Delta w(k+1)\end{aligned} \tag{64}$$

Define

$$\begin{aligned}\gamma(k)=&\varepsilon_1 u(k)\Delta y(k)+\cdots+\varepsilon_{n_y+1}u(k)\Delta y(k-n_y)+\varepsilon_{Ly+1}\Delta u(k)\\&+\cdots+\varepsilon_{Ly+n_u+1}u(k)\Delta u(k-n_u)\end{aligned} \tag{65}$$

We consider the following equation with the vector $\boldsymbol{\eta}(k)$ for each time $k$:

$$\gamma(k)=\boldsymbol{\eta}^T(k)\Delta\boldsymbol{H}(k) \tag{66}$$

Owing to $\|\Delta\boldsymbol{H}(k)\|\neq 0$, (65) must have at least one solution $\boldsymbol{\eta}^*(k)$. Let

$$\begin{aligned}\boldsymbol{\phi}_L(k)=&\boldsymbol{\eta}^*(k)+[\frac{\partial f(\boldsymbol{\varphi}(k-1))}{\partial y(k-1)},\cdots,\frac{\partial f(\boldsymbol{\varphi}(k-1))}{\partial y(k-n_y-1)},0,\cdots,0\\&\frac{\partial f(\boldsymbol{\varphi}(k-1))}{\partial u(k-1)},\cdots,\frac{\partial f(\boldsymbol{\varphi}(k-1))}{\partial u(k-n_u-1)},0,\cdots,0]^T\end{aligned} \tag{67}$$

Then (64) can be rewritten as (57).

Case 4: $L_y\geq n_y+1$ and $1\leq L_u<n_u+1$; $0\leq L_y<n_y+1$ and $L_u\geq n_u+1$.

The proof of Case 4 is similar to the above analysis process; we omit it.

We finished the proof of *Theorem 1*.